\documentstyle[11pt,newpasp,twoside,epsf]{article}
\begin{document}
\title{Fabry-Perot observations at the 6m telescope}
\author{ A.V. Moiseev, V.L. Afanasiev, and S.N. Dodonov}
 \affil{ Special Astrophysical Observatory, Nizhnij Arkhyz,
 369167, Russia }

\begin{abstract}
A current status  of the observations with Interferometer Fabry-Perot (IFP) at the 6m telescope are
described. A reduction of the data obtaining with a new focal reducer SCORPIO in IFP mode are outlined.
\end{abstract}

\section{  SCORPIO design}

The 2D spectroscopy observations with  IFP at the 6m telescope using the CIGALE system were begun  in
early 80s  by teams from Marseille Observatory (France) and from  SAO (Boulesteix  et al., 1982; Amram
et al., 1992). In 1997 a CCD   was attached on the old  focal reducer instead of photon counter. A new
multi-mode focal reducer SCORPIO  (Spectral Camera with Optical Reducer for Photometrical and
Interferometrical Observations) was developed in the SAO RAS in 2000 (Afanasiev et al., this
Proceeding).  Here we describe only
 Fabry-Perot mode of the SCORPIO. For more details see our WWW-site \\
\verb*"http://www.sao.ru/~moisav/scorpio/scorpio.html".

The optics of SCORPIO compensates the  aberrations of the main telescope mirror and provide a total
focal ration $F/2.9$.  The Queensgate IFPs ET-50 installs  into the 35mm collimated beam. SCORPIO are
under full remote control from  a PC. A total DQE (telescope$+$filter$+$IFP$+$CCD) is about $20\%$ in
the $H_\alpha$ spectral region. SCORPIO equipped  with narrow filters ($FWHM=10-15$\,\AA) for IFP
observations in the $H_\alpha$ line for object's velocities $-200...+10.000\,\mbox{km}\,\mbox{s}^{-1}$.

\section{ Calibration and data reduction}

Focal reducer mounted  on the prime focus universal adapter which   has two movable probes for offset
guiding and continuum and spectral lamps  for calibration. A He-Ne-Ar lamp and continuum spectrum lamp
mounted in the integration sphere which produced a uniform lighting. The optics of calibration system
forms the $F/4$ beam on the reducer entrance. There is no systematic shift in velocities because
telecentrical beams used for wavelength calibration.

A CCD has a large advantage in quantum effeciency   in comparison with a  photon counter. However a
readout noise and a reading time do not allow to do very short exposures of the object channels.
Therefore exist a problem of the photometry correction of flux and seeing in the individual channels
because only 1-2 scanning cycles per channel may be accepted on CCD observations, we should take in
account a time dependence of a sky brightness and of an atmospheric extinction. Therefore wavelength
object cube contains a set of narrow rings (Fig.~1c). This  pattern   limits  the velocities
measurements for the low brightness objects. We suggest a method to resolve  this problem. Main idea is
removing the sky counts {\it before} cube linerization. After standard CCD-reduction a mean radial
profile of the sky lines is constructed by  the azimuthal averaging  of the interferometrical rings
(Fig.~1a).   The sky radial profile are subtracted from the object. Fig.~1b shows example of the sky
subtracted images. Variations of an atmospheric extinction, seeing and channel offset shifts could be
measured from the photometry of background stars  and corrected in the cube. Unfortunately our method
requires that about half of field of view will be free from the object emission, but it allows to
obtain a good results in the object cube.

\begin{figure}
\plotfiddle{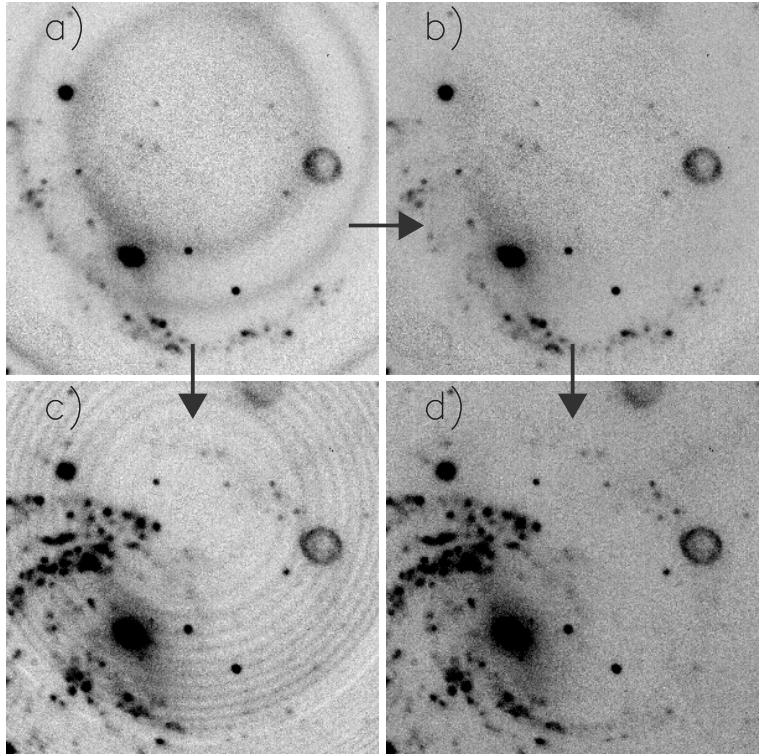}{9.5 cm}{0}{100}{100}{-160}{0} \caption{SCORPIO observations of
NGC~6951 in $H_\alpha$: raw IFP channel with night sky lines ({\bf a}) and after night sky model
removing ({\bf b}). The wavelength cube channel were obtained from cube with sky lines ({\bf c}) and
with night sky lines removed -- {\bf d}. }
\end{figure}


\begin{references}

\reference Amram P., Marcelin M., Bonnarel F., Boulesteix J., Afanas'ev
V.L., Dodonov S.N., 1992, A\&A, 263, 69

\reference Boulesteix J.,  Georgelin Y., Marcelin M., Fort J.A., 1982,
  in ''Instrumentation for astronomy with large optical telescopes'',  223

\end{references}
\end{document}